# A beam monitor detector based on doped silica and optical fibres

S. Braccini[a]*, A. Ereditato[a], F. Giacoppo[a], I. Kreslo[a], K.P. Nesteruk[a#], M. Nirkko[a], M. Weber[a], P. Scampoli[b,c], M. Neff[d], S. Pilz[d] and V. Romano[d]

[a] *Albert Einstein Center for Fundamental Physics,*
*Laboratory for High Energy Physics, University of Bern,*
*Sidlerstrasse 5, CH-3012 Bern, Switzerland*
*E-mail:* `Saverio.Braccini@lhep.unibe.ch`

[b] *Dipartimento di Scienze Fisiche, Università di Napoli Federico II,*
*Complesso Universitario di Monte S. Angelo, I-80126 Napoli, Italy*

[c] *Department of Radiation Oncology,*
*University of Bern, Inselspital, CH-3010 Bern, Switzerland*

[d] *Institute of Applied Physics, University of Bern,*
*Sidlerstrasse 5, CH-3012 Bern, Switzerland*

ABSTRACT: A beam monitor detector prototype based on doped silica fibres coupled to optical fibres has been designed, constructed and tested, mainly for accelerators used in medical applications. Scintillation light produced by Ce and Sb doped silica fibres moving across the beam has been measured, giving information on beam position, shape and intensity. Mostly based on commercial components, the detector is easy to install, to operate and no electronic components are located near the beam. Tests have been performed with a 2 MeV proton pulsed beam at an average current of 0.8 µA. The response characteristics of Sb doped silica fibres have been studied for the first time.

---

[*] Corresponding author.
[#] Also at Faculty of Physics, University of Warsaw, ul. Hoża 69, PL-00681 Warszawa, Poland.

## Contents

## 1. Introduction

Beam monitoring is essential in any particle accelerator and diagnostics devices are usually installed near critical locations, such as beam injection or extraction points [1]. The precise and continuous control of the position, intensity and shape of the beams is crucial in medical applications where ion beams are used to irradiate tumours with millimetre precision or to produce large quantities of radioisotopes for diagnostics and therapy. In the recent years, a remarkable scientific and technological progress led to innovative instruments and to the construction of many accelerator based facilities in hospitals and research centres, mainly dedicated to hadrontherapy [2] and radioisotope production [3].

    In the case of hadrontherapy, proton or carbon ion beams with currents of the order of 1 nA have to be controlled during acceleration, transport and extraction to be able to deliver the dose distribution prescribed by the treatment plan to the patient. For the production of radioisotopes, proton beams of high intensity, in the range (10-500) µA, are used and the control of the target bombardment is very important for an efficient, safe and reliable production. Although the beam intensities are quite different, the problem of the control of the beams is important in both cases and the use of reliable, robust and easy to use detectors is fundamental for debugging, validation, quality control and routine operations. Different beam monitoring systems are employed in accelerators for medical applications [4], mostly based on technologies developed in nuclear and particle physics. It has to be remarked that most of the commonly used detectors – such as Faraday cups – are destructive and their use is not possible in parallel with irradiation. Some nearly non-destructive detectors have been developed for hadrontherapy (see for example Ref. [5], [6] and [7]), often based on sophisticated and expensive technologies using radiation sensitive front-end electronics.

    Due to the cost, size and complexity of beam monitor detectors, the number of installed devices is often minimum with direct implications on the optimization procedures. In many cases, the measurement of the beam is performed by means of films and other passive devices. In case of radioisotope production facilities, this implies a time-consuming procedure due to the

several opening and closing of the bunker doors and to the appreciable radiation dose to the operators, avoidable by using remotely controlled detectors. A typical ambient dose level allowing the access to a cyclotron bunker could be up to 50 µSv/h.

In this paper, we report on a robust and easy to use beam monitoring detector based on doped silica fibres moving across the beam. This device has been principally developed for the external beam line of the Bern cyclotron laboratory [8]. This facility, equipped with an 18 MeV proton cyclotron (maximum current 150 µA) for the production of radioisotopes and research, will be operational in 2012. The proposed general-purpose system is able to operate with beam intensities ranging from single particle to several µA, to measure the beam position and to operate in high intensity radiation fields due to the absence of any radiation sensitive front-end electronics. Since this device is very thin, it can be installed along the beam transport lines.

## 2. Rationale for a beam monitor detector based on doped silica and optical fibres

The detection of charged particles by means of scintillation is a very well known process on which many commonly used particle detectors are based. In particular, pure silica and plastic scintillating fibres are used for specific applications in nuclear and particle physics [9][10].

For beam monitoring, plastic fibres are appropriate only for low currents due to their very low resistance to heat. In hadrontherapy, an array of plastic scintillating fibres read out by a CCD has been recently developed for the Italian National Centre for Hadrontherapy CNAO [11]. For high currents, $Ce^{3+}$ doped quartz surfaces have been used [12][13] and resistance up to temperatures of 1300°C has been reported. Scintillation light emitted by $Ce^{3+}$ doped quartz has been measured to be peaked at 395 nm, making this material suitable for a read-out based on photomultipliers (PMT). $Ce^{3+}$ doped silica fibres have been used for beam monitoring [14][15][16], neutron detection [17] and, recently, for dosimetry in hadrontherapy [18].

Due to their scintillation properties and their resistance to heat, uncoated doped silica fibres represent a promising instrument for the monitoring of beams ranging from fractions of nA to several µA. Contrary to inductive detectors, a sensor based on fast scintillation is easily adaptable either to pulsed or continuous beams. It has to be noted that fibres arranged in an array are naturally subject to non-uniform irradiation and their relative response changes with time due to the damage induced by radiation, thus requiring a frequent calibration procedure. Moreover, in the case of radioisotope production, the intense radiation fields, mainly due to neutrons, do not allow the use of front-end electronics or CCD readout.

In this framework, we developed the device presented in Fig. 1, based on a single doped silica fibre moving transversally through the beam. Since uncoated doped silica fibres are very fragile, difficult to handle and have a considerable attenuation length (of the order of several dB/m) a sensor fibre of about 10 cm in length is recommended. To transport the optical signal produced by the scintillation induced by the impinging beam particles, the doped fibre is coupled to a commercial optical fibre by means of a connection. In this way, the signal can be transported over several meters distance with negligible attenuation (of the order of tens dB/km). The connection requires some care in order to limit light losses. The connection is located inside the vacuum chamber and the commercial optical fibre is extracted into air with a feedthrough. A remotely controlled linear translation stage coupled to a vacuum-tight feedthrough or a bellow can be used to move the sensing fibre across the beam.

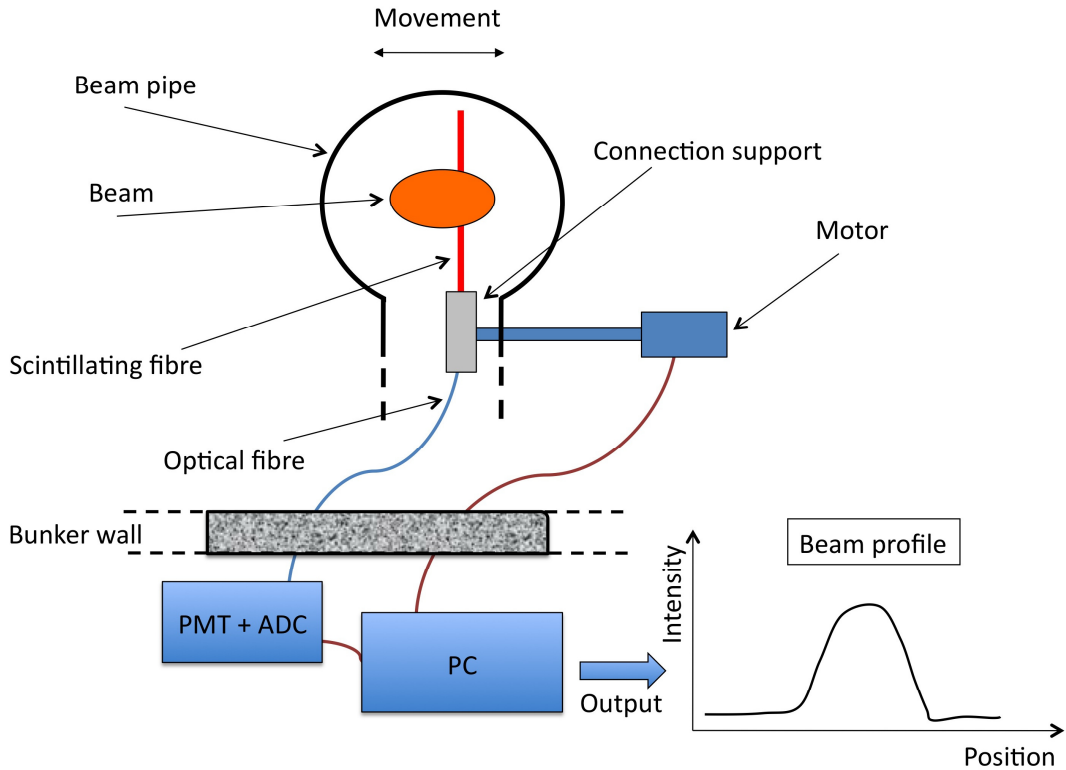

**Figure 1.** Principle and general scheme of a beam monitor detector based on a single moving scintillating doped silica fibre coupled to an optical fibre.

## 3. Materials and methods

A first prototype detector has been designed, constructed and tested. Cerium and antimony doped silica sensing scintillating fibres have been used. In view of the installation at the Bern cyclotron laboratory, beam tests have been performed with the 2 MeV RFQ linac installed at the Laboratory for High Energy Physics (LHEP) of the University of Bern.

### 3.1 The RFQ accelerator

The 2 MeV RFQ linac was originally conceived for the calibration of the electromagnetic calorimeter of the L3 experiment at LEP [19][20][21]. Several years after LEP was dismantled, it has been refurbished and put back into operation in Bern. This accelerator is capable of producing H$^-$ beams with spills at a repetition rate in the range (10 – 150) Hz. The peak current and the beam pulse width can be adjusted in the range (1 – 5) mA and (1 – 25) μs, respectively. The nominal emittance of the H$^-$ beam is 0.5 mm mrad. Beam cross sections from 2×2 mm$^2$ to 50×50 mm$^2$ can be obtained in the extraction beam line. The beam tests presented in this paper were performed at a position along the beam axis corresponding to a beam size of about 20 mm diameter. This value is close to beam sizes used in radioisotope production cyclotrons.

### 3.2 The first prototype detector

The first prototype detector is shown in Figs. 2 and 3. Aiming at realizing a compact, cost effective and easy to use system, commercial components have been used whenever possible.

The main body of the detector is an ISO KF-40 cross joint on which a vacuum tight linear motion feedthrough (Pfeiffer Vacuum DS040A) is mounted (Fig. 3). This system is characterized by very limited leaks, reported by the manufacturer to be less than $10^{-9}$ mbar l/s in static situation and less than $10^{-4}$ mbar l/s during movement. A 5 cm long aluminium support is screwed to the linear motion feedthrough to hold the sensing fibre and to connect it to the commercial optical fibre.

For this first prototype, the movement is manual, while a motorized system is presently under development. To accurately measure the position of the sensing fibre when moving through the beam, calibrated blocks are used.

Inside the mechanical support, a V-groove is milled along the 5 cm to accommodate fibres with a maximum diameter of 600 μm. In this way, the two fibres are mechanically put in contact and fixed by means of an aluminium plate screwed on the top to assure rigidity to the system. Contrary to fusion splicing, the use of a V-groove connection allows for an easy replacement of the sensing fibre. With this mechanical configuration light losses in the range of 20%-50% have been measured. Even if the reproducibility has still to be optimized, the light losses are constant during the operation of the device.

To extract the optical signal into air, a vacuum tight feedthrough made by a blank flange with a hole sealed with epoxy glue is used (Fig. 3). Leak tests using a helium spectrometer (Pfeiffer Vacuum SmartTest HLT 570) have shown good vacuum tightness within the sensitivity of the instrument.

To assure sufficient flexibility to the optical fibre during the movement, a 90° elbow joint is mounted between the bottom of the cross joint and the feedthrough (Fig. 3).

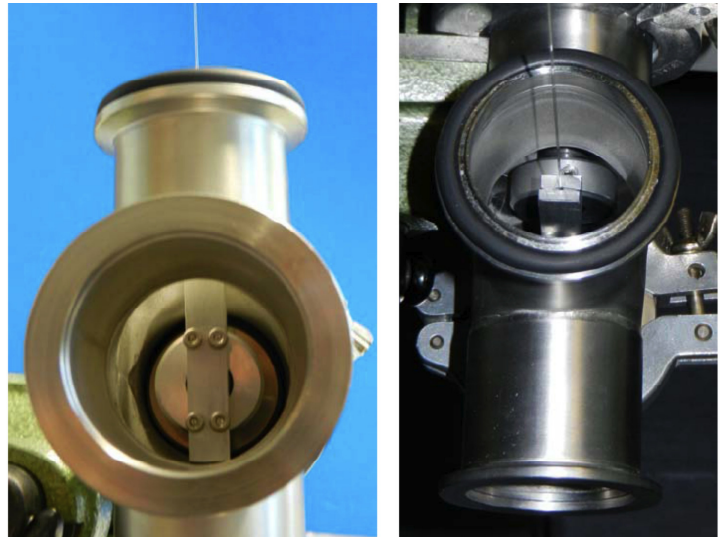

**Figure 2.** Details of the construction of the first prototype detector. The side view (left) shows the V-groove moving support where the doped and the optical fibres are connected. The sensing fibre is visible on the top. The top view (right) shows the moving support connected to the final part of the linear motion feedthrough and the doped sensing fibre. The beam direction is orthogonal to the movement of the fibre.

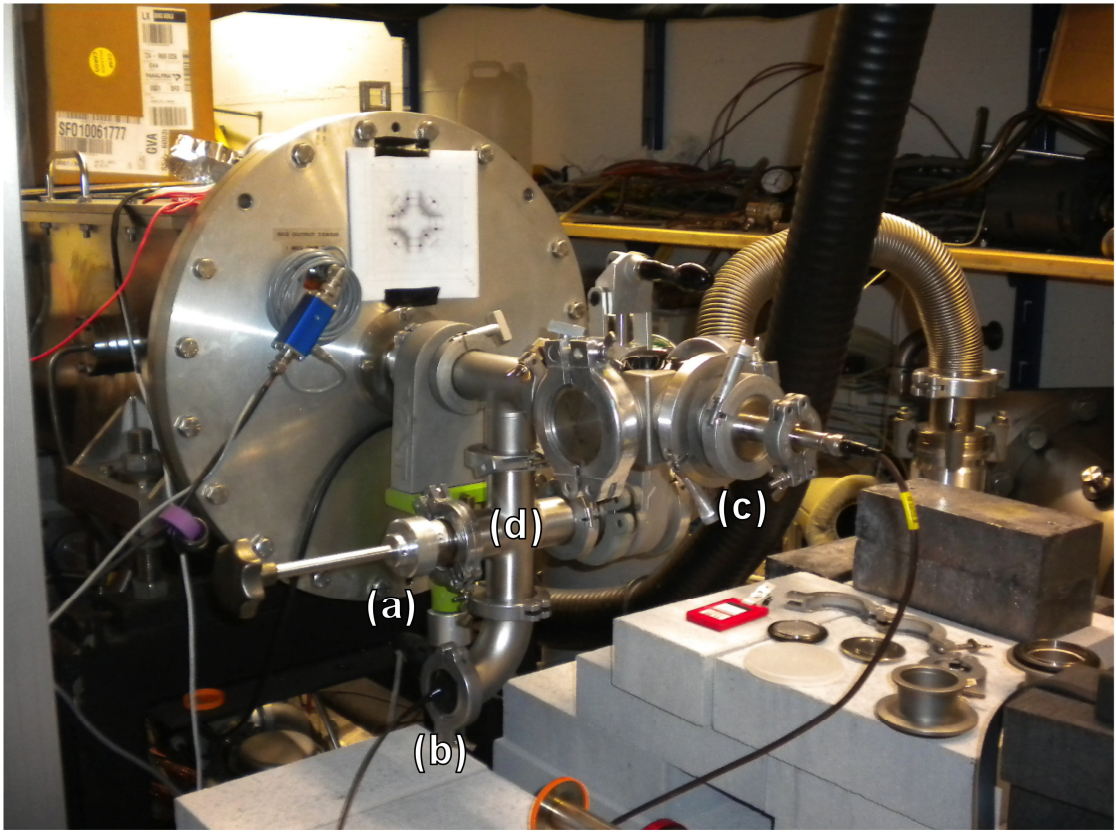

**Figure 3.** The first prototype detector installed on the RFQ accelerator at LHEP. The linear motion (a) and the optical fibre feedthrough (b) are visible together with the Faraday cup (c) used for measuring the total intensity of each spill. The V-groove connection supporting the sensing fibre is contained inside the cross joint (d).

For read-out and digitization, a photomultiplier (Philips 56AVP B2) and a peak sensing ADC (LeCroy 2259B) have been used. During the beam tests, a Faraday cup located downstream the sensing fibre has been used. Both signals, from the PMT and the Faraday cup, are digitized by the peak sensing ADC. The signal from the Faraday cup has been used for the ADC gate.

A 20-m long multimode optical fibre (ThorLabs BFH48-400) was used to transport light to the PMT in the counting room (see Table 1).

**Table 1.** Main characteristics of the optical fibre used for transporting the signal.

| Core/cladding materials | Pure silica/hard polymer |
|---|---|
| Core diameter ($\mu$m) | $400 \pm 8$ |
| Cladding diameter ($\mu$m) | $430 \pm 9$ |
| Numerical aperture (NA) | $0.48 \pm 0.02$ |
| Attenuation at 500 nm (dB/km) | 30 |

### 3.3 Doped silica scintillating fibres

$Ce^{3+}$ and $Sb^{3+}$ doped fibres have been considered for the prototype presented in this paper. The $Ce^{3+}$ doped fibres were industrially produced several years ago and are no more commercially available. They were used for the detector described in Ref. [14]. The $Sb^{3+}$ doped fibres have been produced by the Institute of Applied Physics (IAP) of the University of Bern [22][23]. The main characteristics of the selected fibres are reported in Table 2.

**Table 2.** Main characteristics of the selected doped silica fibres.

| Dopant | Composition | Core diameter (μm) | Cladding diameter (μm) | Numerical aperture (NA) | Length (mm) | Commercial name |
|---|---|---|---|---|---|---|
| $Ce^{3+}$ | $SiO_2$ (55 %), MgO (24 %), $Al_2O_3$ (11 %), $LiO_2$ (6 %), $Ce_2O_3$ (4 %) [17] | Unknown | 520 ± 20 | 0.25 | 120 | NE901 |
| $Sb^{3+}$ | (0.5%) $Sb^{3+}$:$SiO_2$ [23] | 461 ± 16 | 570 ± 20 | 0.05 | 120 | - |

While the scintillation properties of $Ce^{3+}$ doped materials are known [12][13][24][25][26][27][28][29], the response to ionizing radiation of $Sb^{3+}$ doped fibres has never been studied.

To assess the light yield, tests have been performed with a collimated 370 MBq $^{90}$Sr beta source located about 1 cm away from the sensing fibre. Being the average energy of the emitted electrons 1.13 MeV, corresponding to a range in glass of 2.2 mm, most of the electrons cross the fibre completely. The maximum deposited energy is estimated to be 180 keV. By measuring the pulse height of the signal given by the PMT and taking into account its gain, quantum efficiency and the total collection efficiency, the light yield is estimated to be 0.6% and 0.4% of the deposited energy for $Ce^{3+}$ and $Sb^{3+}$ doped fibres, respectively. The result obtained for the cerium doped fibres is in good agreement with Refs. [13] and [29].

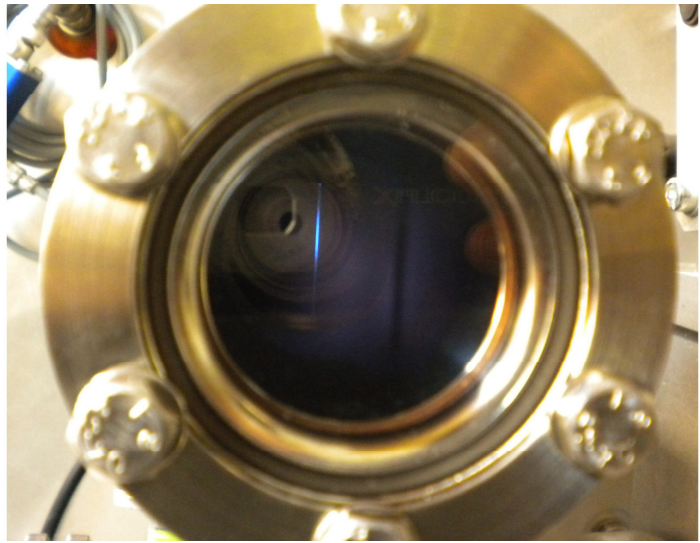

**Figure 4.** The 570 μm diameter $Sb^{3+}$ doped fibre irradiated by the 2 MeV proton beam. The emitted blue light clearly shows the beam profile. Since protons are stopped by the fibre, its shadow is visible as a black line in the glow light produced by the beam hitting the glass window.

To further investigate the characteristics of the $Sb^{3+}$ doped fibres, measurements have been performed with the 2 MeV proton beam produced by the RFQ linac. The 2 MeV protons loose all their energy in the fibre since they range out in 45 μm of glass. When irradiated by the beam, the fibre glows, emitting a very clear blue light signal, as shown in Fig. 4.

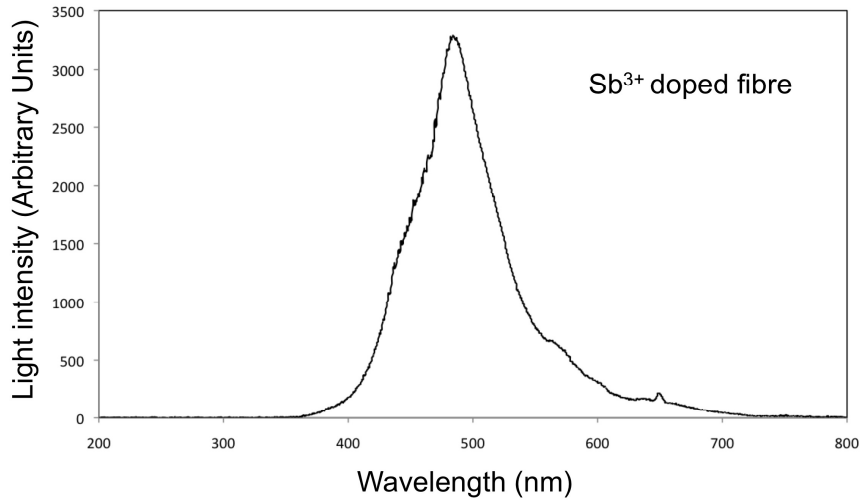

**Figure 5.** Measured spectrum of the wavelengths emitted by a $Sb^{3+}$ doped fibre irradiated with 2 MeV accelerated protons.

The spectrum of the emitted light (Fig. 5) has been analyzed by means of a grating based spectrometer (Avantes AVS-USB2000, range 190 nm – 860 nm) connected after the 20 m optical fibre. The maximum of the emission is located around 490 nm. This feature allows for good quantum efficiency when coupled to a PMT. Other structures are visible, probably due to transitions of antimony atoms in the silica glass. In Ref. [23] a maximum intensity around 630 nm has been reported when the fibre is excited with 337 nm and 488 nm from nitrogen and argon lasers, respectively. The difference in the emitted wavelengths is due to the completely different excitation mechanisms of lasers and accelerated proton beams.

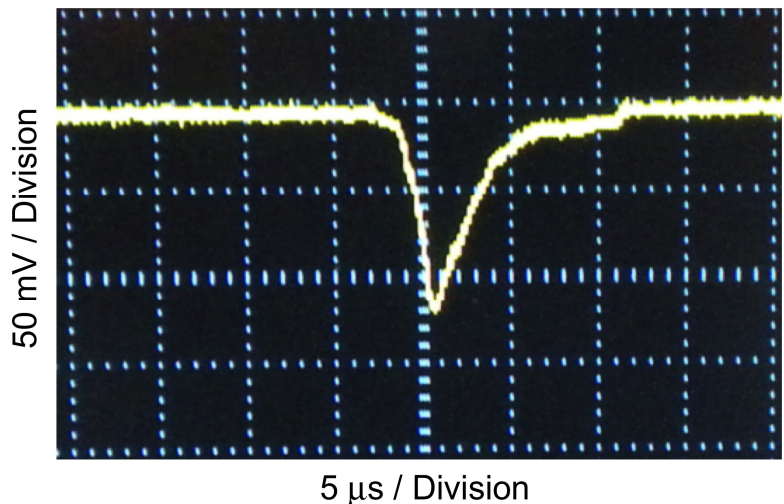

**Figure 6.** Typical single signal produced by a $Sb^{3+}$ doped fibre during a proton spill. The nominal duration of the spill is set to about 3 μs.

A typical signal from the PMT measured with an oscilloscope (Tektronix TDS 2024C, 200 MHz bandwidth) is shown in Fig. 6. The drop of the signal after about 10 μs is due to the accelerator since the same temporal behaviour has been measured by the Faraday cup. For these measurements, the beam intensity was set to about $10^{11}$ protons per spill, which, at a repetition rate of 50 Hz, leads to an average total current of about 0.8 μA. In these conditions about $10^{10}$ protons hit the fibre when centred on the beam. With a bias voltage of 1600 V, the signal is evaluated to be produced by about 37500 photoelectrons, and taking into account the quantum efficiency of 12% at 490 nm by about 315000 photons reaching the photocathode through the 20 m optical fibre.

The emission of scintillation light depends on the temperature of the scintillator. Since high beam intensities produce a considerable heat release, this phenomenon has a significant influence on the output signal. To investigate this effect with the $Sb^{3+}$ doped fibre, specific irradiations have been performed.

The outputs from the PMT and from the Faraday cup have been recorded with and without the beam. The average values without the beam (pedestals) have been subtracted from the ADC measurements with the beam to obtain the corresponding signals. The ratio of the signal from the sensing fibre with respect to the Faraday cup has then been calculated to correct for effects due to fluctuations of the beam intensity. The result is shown in Fig. 7 for 5000 consecutive spills which, at the repetition rate of 50 Hz, correspond to 100 seconds. The total charge hitting the fibre is estimated to be 2.7 μC with a total corresponding delivered dose on the volume of the fibre of 560 kGy. This measurement shows a decrease of the light output of about a factor five when long irradiations are performed.

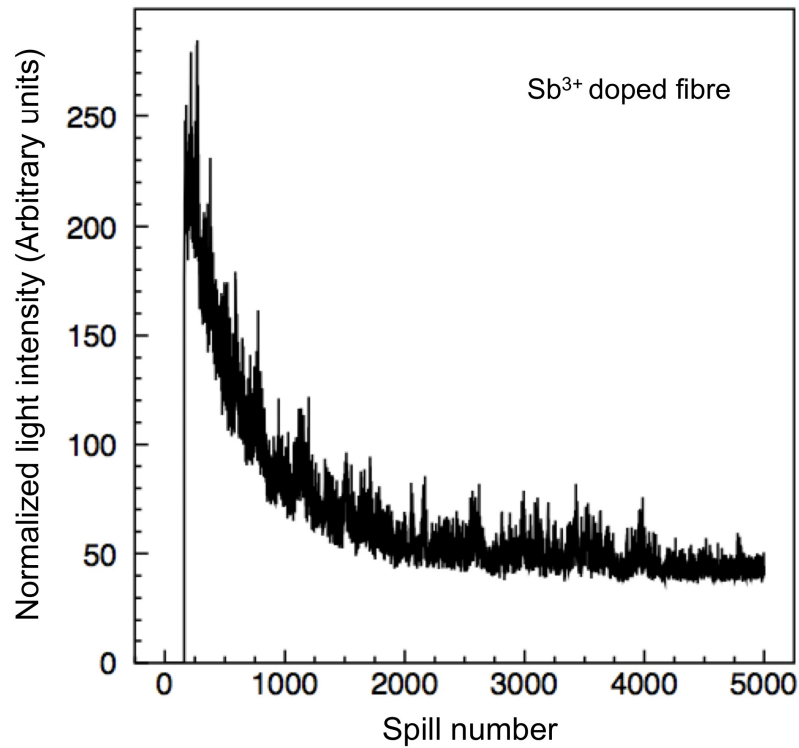

**Figure 7.** Light output for 5000 successive spills of equal intensity. The effect of heating leads to a decrease of the light output of about a factor five. The repetition rate of the accelerator is 50 Hz.

With the device presented in this paper, this effect leads to distortions of the beam profile since the fibre is moved manually through the beam and different temperatures correspond to different positions. However, using a motorized device which allows sweeping the beam in less than one second, this effect will be reduced since the light loss is estimated to be about 4% in one second for a cold fibre.

After irradiation, we observed that the light output returns at the initial value if enough time is allowed for cooling down the fibre to room temperature. Irradiations with up to 50000 spills – corresponding to 1000 seconds – have been performed and no effects due to permanent radiation damage have been observed. Further investigations on radiation hardness are foreseen using the external beam line of the 18 MeV cyclotron in Bern.

## 4. Beam profile measurements

The horizontal beam profile of the 2 MeV pulsed proton beam from the RFQ accelerator has been measured using the prototype detector described in the previous Section and installed in the extraction beam line, as shown in Fig. 3.

The beam profile has been measured by moving the fibre across the beam in steps of 1 mm and by acquiring the peak of the signal for 1000 consecutive spills after stabilization of the light output with respect to temperature in each position. The centre of the beam pipe corresponds to the position of the fibre at 14 mm.

Cerium and antimony doped fibres have been used and, as an example, ADC spectra for two different positions are reported in Fig. 8. The set-up of the accelerator was not the same for the two measured profiles therefore the two measured beams are similar but not identical.

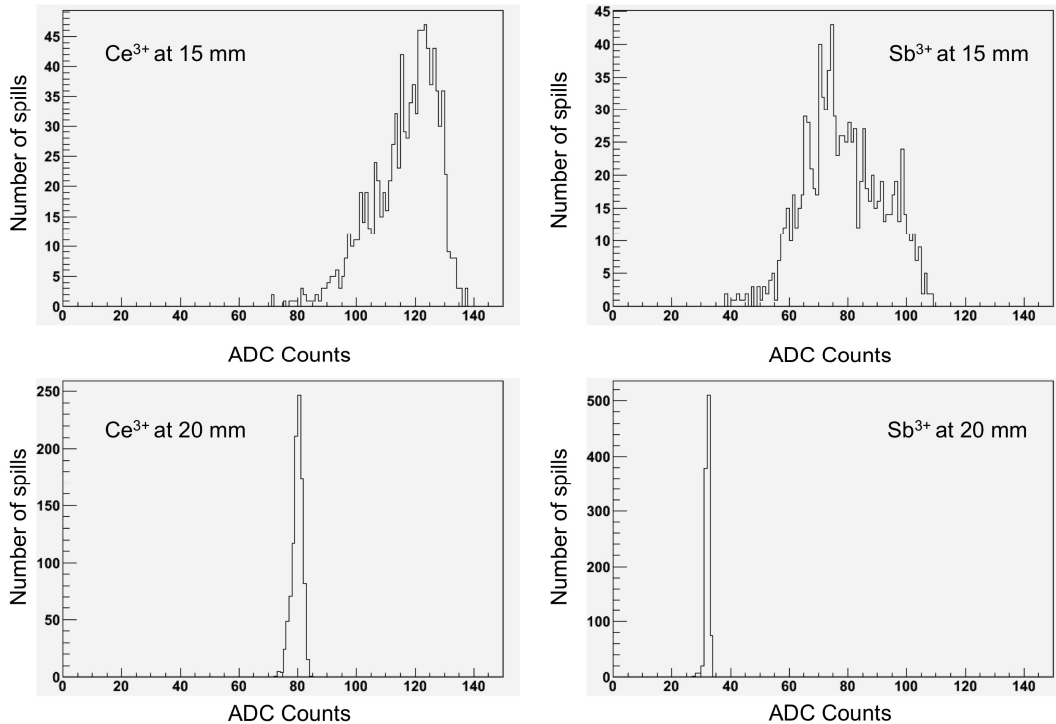

**Figure 8.** ADC spectra acquired for 1000 spills at 15 mm and 20 mm with cerium (left) and antimony (right) doped silica fibres.

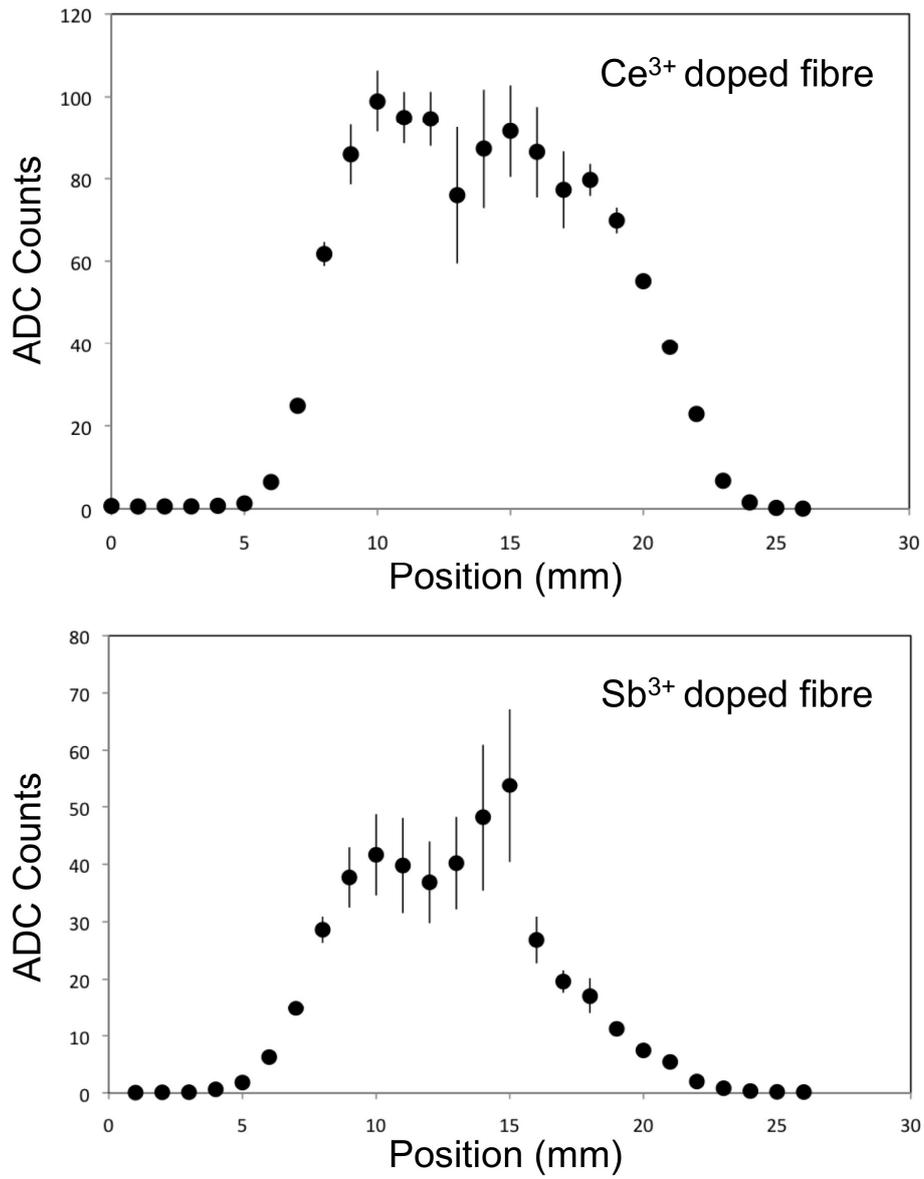

**Figure 9.** Beam profiles measured with the $Ce^{3+}$ (top) and $Sb^{3+}$ (bottom) doped silica fibre moved in steps of 1 mm through the beam. The double structure of the beam is a known effect of the accelerator. The error bars correspond to the rms of the pulse height distributions and are mainly due to beam intensity fluctuations. The two profiles have been measured with different set-ups of the accelerator.

The horizontal beam profiles obtained with $Ce^{3+}$ and $Sb^{3+}$ doped fibres are reported in Fig. 9. For each position of the sensing fibre, the average of the pulse height distribution has been calculated and the average value in absence of the beam (pedestal) subtracted. The error bars correspond to the rms of the ADC output distributions. The major contribution to this error is due to beam intensity fluctuations. No attempt is made to correct for temperature effects on the light output, due to different temperatures in different positions. The difference between the two profiles is mainly due to a different set-up of the accelerator since identical set-ups cannot be reproduced with our linac after shut-downs. A contribution could also be due to a different distortion due to temperature for the two fibres. In both profiles, a double structure of the beam is observed, corresponding to a known feature of the accelerator.

The reproducibility of the measurements has been checked by acquiring twice the pulse height with an oscilloscope in all the positions. All the values are found to be compatible within the uncertainties. Furthermore, two points (at 8 mm and 10 mm) for the cerium and one point (at 10 mm) for the antimony doped fibre have been re-measured with the ADC and found to be within one standard deviation.

As far as the spatial resolution is concerned, the position of the beam can be measured with good precision with the prototype device described in this paper. It has to be remarked that the diameters of the sensing fibres used in this first prototype are 520 and 570 μm for $Ce^{3+}$ and $Sb^{3+}$, respectively, and that the use of thinner fibres will allow a better precision and produce a smaller perturbation to the beam.

## 5. Conclusions and outlook

A beam monitor detector based on doped silica and optical fibres is proposed and a first prototype has been constructed and successfully tested. Beam profiles have been measured using $Ce^{3+}$ and $Sb^{3+}$ doped fibres moved across the beam. Mostly based on commercial components and without sophisticated movement devices and front-end electronics, this detector is capable to measure precisely the position, intensity and shape of low current beams, as those employed in cancer hadrontherapy. For beams of higher intensities, of the order of a few μA, temperature effects lead to distortions of the measured profiles. The device is anyway capable to detect spikes and localize spots of higher intensity, an important bit of information for radioisotope production accelerators.

Cerium and antimony doped silica fibres are suitable instruments for beam profile measurements. $Sb^{3+}$ doped fibres have the advantage of a rapid, cost effective and relatively simple production method using granulated oxides [22][23]. $Ce^{3+}$ doped fibres are no more commercially available and cannot be produced with the above mentioned method.

With respect to beam monitor devices based on array of fibres, a single fibre passing through the beam has the advantage of limited aging effects due to non uniform irradiation. Increasing the speed of the fibre by means of a motorized device, distortion effects due to temperature can be limited. A motorized device based on the prototype presented in this paper, can be employed in large dose rate radiation environments and can be easily installed in beam transport lines.

The tests of this first prototype triggered further developments toward the construction of a motorized two dimensional beam profiler, the test of other doped silica fibres and the study of their radiation hardness. The use of thinner fibres to reduce at minimum the perturbation of the beams will also be investigated. Efforts towards the miniaturization of the system would eventually allow its use in connection with standard targets used for PET radioisotope production.

**Acknowledgments**

We acknowledge contributions from LHEP engineering and technical staff. We would like to thank G. Molinari for providing us with the cerium doped fibres and S. Haug for his constructive comments in reviewing this manuscript. We are greatly indebted to the ICSC-World Laboratory, owner of the 2 MeV proton accelerator, and to its president A. Zichichi for giving us the possibility to operate this device at LHEP.